\documentclass[letterpaper,twocolumn,prl,aps,superscriptaddress,amsmath,amssymb,floatfix]{revtex4-1}
\usepackage{mathptmx}
\DeclareMathAlphabet{\mathcal}{OMS}{cmsy}{m}{n}
\usepackage[latin9]{inputenc}
\usepackage{color}
\usepackage{verbatim}
\usepackage{float}
\usepackage{amsmath}
\usepackage{amssymb}
\usepackage{graphicx}
\usepackage{esint}
\usepackage[unicode=true,
 bookmarks=true,bookmarksnumbered=false,bookmarksopen=false,
 breaklinks=false,pdfborder={0 0 1},backref=false,colorlinks=true]
 {hyperref}
\hypersetup{linkcolor=magenta,urlcolor=blue,citecolor=blue,pdfstartview={FitH},hyperfootnotes=false}

\makeatletter

\usepackage{textcomp}
\usepackage{epstopdf}

\pdfpageheight\paperheight
\pdfpagewidth\paperwidth

\@ifundefined{textcolor}{}{%
 \definecolor{BLACK}{gray}{0}
 \definecolor{WHITE}{gray}{1}
 \definecolor{RED}{rgb}{1,0,0}
 \definecolor{GREEN}{rgb}{0,1,0}
 \definecolor{BLUE}{rgb}{0,0,1}
 \definecolor{CYAN}{cmyk}{1,0,0,0}
 \definecolor{MAGENTA}{cmyk}{0,1,0,0}
 \definecolor{YELLOW}{cmyk}{0,0,1,0}
}

\usepackage{xcolor}
\usepackage{soul}
\definecolor{blue}{rgb}{0,0,1}
\definecolor{red}{rgb}{1,0,0}
\definecolor{green}{rgb}{0,1,0}

\@ifundefined{textcolor}{}{%
 \definecolor{BLACK}{gray}{0}
 \definecolor{WHITE}{gray}{1}
 \definecolor{RED}{rgb}{1,0,0}
 \definecolor{GREEN}{rgb}{0,1,0}
 \definecolor{BLUE}{rgb}{0,0,1}
 \definecolor{CYAN}{cmyk}{1,0,0,0}
 \definecolor{MAGENTA}{cmyk}{0,1,0,0}
 \definecolor{YELLOW}{cmyk}{0,0,1,0}
}

\usepackage{xcolor}\usepackage{soul}
\definecolor{blue}{rgb}{0,0,1}
\definecolor{red}{rgb}{1,0,0}
\definecolor{green}{rgb}{0,1,0}

\usepackage{soul}

\newcommand{\cmt}[1]{{}}

\makeatother

\begin{document}

\preprint{APS/123-QED}

%\title{Backward Stimulated Brillouin Scattering in Suspended Lithium Niobate Nanowaveguides}

\title{On-chip Brillouin Amplifier in Suspended Lithium Niobate Nanowaveguides}% Force line breaks with \\

\author{Simin Yu}
\author{Ruixin Zhou}
\affiliation{School of Information Science and Technology, ShanghaiTech University, Shanghai 201210, China.}%
\author{Guangcanlan Yang}
\affiliation{CAS Key Laboratory of Quantum Information, University of Science and Technology of China, Hefei 230026, China.}
\author{Qiang Zhang}
\affiliation{Quantum Optics and Quantum Optics Devices, Institute of Opto-Electronics, Shanxi University, Taiyuan 030006, China.}%
\author{Huizong Zhu}
\affiliation{School of Information Science and Technology, ShanghaiTech University, Shanghai 201210, China.}%
\author{Yuanhao Yang}
\affiliation{CAS Key Laboratory of Quantum Information, University of Science and Technology of China, Hefei 230026, China.}
\author{Xin-Biao Xu}
\affiliation{CAS Key Laboratory of Quantum Information, University of Science and Technology of China, Hefei 230026, China.}
\author{Baile Chen}
\affiliation{School of Information Science and Technology, ShanghaiTech University, Shanghai 201210, China.}
\author{Chang-Ling Zou}
\email{clzou321@ustc.edu.cn}
\affiliation{CAS Key Laboratory of Quantum Information, University of Science and Technology of China, Hefei 230026, China.}
\author{Juanjuan Lu}
\email{lujj2@shanghaitech.edu.cn}
\affiliation{School of Information Science and Technology, ShanghaiTech University, Shanghai 201210, China.}

\date{\today}% It is always \today, today,

\begin{abstract}
%Thin film lithium niobate (TFLN) has emerged as a promising material platform for integrated nonlinear photonics, enabling applications such as broadband Kerr soliton microcombs and high-speed electro-optic modulation. While stimulated Brillouin scattering has been numerically proposed in TFLN, achieving sufficient gain remains challenging due to the need for both low optical and mechanical losses of the device. By simultaneously optimizing the confinement of both photonic and phononic modes in a membrane-suspended lithium niobate nanowaveguides, we have demonstrated a record-low net-gain threshold of 2.83\,mW for a Brillouin amplifier. At a pump power of 53.5\,mW, we achieved a Brillouin gain of 0.61\,dB, corresponding to a gain coefficient of 112.0\,m$^{-1}$W$^{-1}$. Furthermore, we demonstrate Brillouin frequency tuning by varying either the pump frequency or chip temperature. Our work not only validates the feasibility of achieving strong guided Brillouin interaction using suspended TFLN nanowaveguides but also paves the way for novel on-chip sensing and signal processing applications.

Thin film lithium niobate (TFLN) has emerged as a leading material platform for integrated nonlinear photonics, enabling transformative applications such as broadband Kerr soliton microcomb and high-speed electro-optic modulation. While stimulated Brillouin scattering has been numerically proposed in TFLN, achieving sufficient gain remains challenging due to the requirement for the simultaneous low optical and mechanical losses of the device. In this work, we systematically characterize the angle-dependence of Brillouin gain coefficients in x-cut membrane-suspended TFLN nanowaveguides, taking into account the anisotropy of the photoelastic coefficients in lithium niobate. We report a Brillouin gain coefficient of 129.5\,m$^{-1}$W$^{-1}$ and further demonstrate the Brillouin frequency tuning through variations in either pump frequency or chip operating temperature. Based on the suspended TFLN nanowaveguide, by optimizing the confinement of both photonic and phononic modes, we have achieved a Brillouin amplifier with a record-high gain of 8.5\,dB. This result not only validates the feasibility of strong guided Brillouin interaction using suspended TFLN nanowaveguides, but also paves the way for novel on-chip sensing and signal processing applications.
\end{abstract}

\maketitle

%\tableofcontents

\section{\label{sec:intro}Introduction}
Stimulated Brillouin scattering (SBS), a third-order optical nonlinear process, involves the stimulated interaction between optical photons and acoustic phonons. First demonstrated in 1964~\cite{chiao1964}, SBS has since found widespread use in various fiber optical applications, including coherent light generation~\cite{otterstrom2018,Gundavarapu2019}, optical sensing~\cite{Kobyakov:10,Pang:20}, narrow-linewidth optical filters~\cite{marpaung2015,Choudhary:16}, and signal processing~\cite{shao2019,sarabalis2021,mayor2021}. Beyond its role in traditional optical communication and sensing applications, SBS has also profoundly impacted interdisciplinary areas such as coherent phonon generation~\cite{PhysRevLett.55.2152,PhysRevLett.53.989}, synchronization of mechanical oscillators~\cite{Han:14,PhysRevLett.109.233906,PhysRevLett.123.017402}, ground state cooling~\cite{PhysRevLett.132.023603,chenBrillouinCoolingLinear2016,bahlObservationSpontaneousBrillouin2012}, squeezed light generation~\cite{PhysRevA.33.4008,Li:22}, and microwave-optical conversion~\cite{Yang2024b}.

Taking advantage of their small mode volumes, many micro- and nano-scale structures have been employed to enhance the photon-phonon interaction~\cite{liNanophotonicCavityOptomechanics2015,Wiederhecker2019}, such as sub-wavelength diameter fibers~\cite{florez2016brillouin,Zeng:22}, photonic nanowires~\cite{vanlaer2015}, hybrid plasmonic waveguides~\cite{linStrongOptomechanicalInteraction2015,fangPlasmonicCouplingBow2011}, and photonic integrated circuits (PICs)~\cite{Safavi-Naeini:19,lin2024opticalmultibeamsteeringcommunication}. Among these, PICs are particularly appealing due to its advantages in terms of device miniaturization and great scalability. So far, various nanophotonic material platforms have been investigated for on-chip SBS, involving silicon~\cite{shin2013,Otterstrom2019,Kittlaus2017}, chalcogenide~\cite{Pant:11,https://doi.org/10.1002/adfm.202105230,yuanEnhancingBrillouinScattering}, silicon nitride~\cite{gyger2020,Botter2022,10.1063/5.0178804}, and aluminum nitride~\cite{LiOE:22}. Most recently, thin film lithium niobate (TFLN) stands out as a promising nonlinear photonic platform, offering high piezoelectric~\cite{jiang2020a}, electro-optic~\cite{wangIntegratedLithiumNiobate2018}, and photoelastic coefficients~\cite{weis1985}, along with high integration potential~\cite{fengIntegratedLithiumNiobate2024,zhuIntegratedPhotonicsThinfilm2021}. High performance acousto-optical devices based on TFLN have been demonstrated ~\cite{shaoPhononicBandStructure2019,wan2022}. Moreover, SBS has been observed in unsuspended lithium niobate nanowaveguides on either silica~\cite{RodriguesCLEO:24,YeCLEO:24} or sapphire~\cite{Yang2024}. The potential of the TFLN platform for SBS, however, is far from being unleashed, as the Brillouin gain coefficients demonstrated in recent works remain much lower than the theoretical prediction~\cite{JOSAB2023}.

In this work, we present the first experimental demonstration of backward SBS in suspended TFLN nanowaveguides. This design significantly enhances the confinement of acoustic waves, thereby strengthening the acousto-optic interaction. We investigate backward SBS in x-cut TFLN suspended waveguides with various rotation angles ($\theta$) relative to its crystalline z-axis. A maximum Brillouin gain coefficient of 129.5\,m$^{-1}$W$^{-1}$ is achieved at the angle of 135\,$^\circ$. We also explore the tuning of Brillouin frequency by adjusting the pump frequency and chip operating temperature, realizing the tuning rates of $-7.00\pm0.10$\,MHz/nm and $-0.40\pm0.02$\,MHz/$^\circ\mathrm{C}$, respectively. Our experiments demonstrate a record-high Brillouin gain of 8.5\,dB for the Brillouin amplifier. This work not only validates the feasibility of strong guided Brillouin interaction using suspended TFLN nanowaveguides but also lays the foundation for advanced on-chip sensing and signal processing applications.

\begin{figure*}
\includegraphics[width=1.0\textwidth]{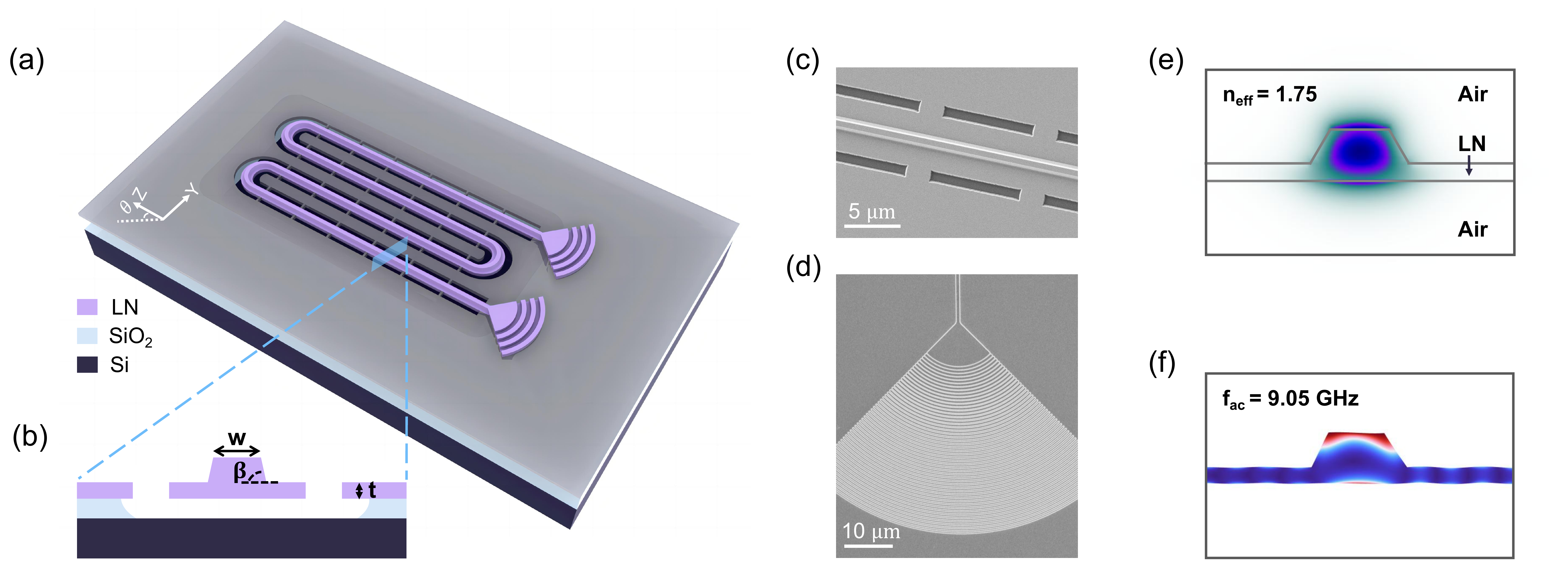}
\caption{\label{devicedesign} \textbf{Suspended lithium niobate nanowaveguide for enhanced Brillouin interaction.} (a) Schematic of the meander-shaped suspended lithium niobate (x-cut) waveguides with a rotation angle of 0\,$^\circ$ with respect to the crystalline $z$-axis. (b) Cross-sectional view of the active photon-phonon interaction region. (c)-(d) Scanning electron microscope (SEM) images of suspended region of the fabricated devices for Brillouin interaction (c) and the self-imaging apodized grating coupler for efficient light coupling (d). (e)-(f) Simulated model profiles of tightly confined transverse electric-like (TE-like) guided optical mode at $\lambda$ = 1550\,nm and phonon mode at 9.05\,GHz.}
\end{figure*}

\section{Device Design and Fabrication}
Figure\,\ref{devicedesign}(a) illustrates the schematic of a suspended LN device designed for enhanced Brillouin interaction. The meander-shaped lithium niobate waveguides are fabricated on a 600\,nm thick x-cut undoped congruent lithium niobate on insulator (LNOI) wafer (NanoLN Electronics). The presence of the silica substrate beneath the LN layer can significantly influence the guided acoustic wave in the LN waveguides, due to the weak confinement of acoustic wave from the small velocity contrast between the LN and silica and the strong acoustic damping in the silica at around 10\,GHz. To mitigate these effects and optimize the confinement of both optical and acoustic waves, we propose a suspended structure that isolates the LN waveguide structure from the silica layer. This suspended design not only enhances the acoustic confinement but also ensures excellent overlaps between the optical and acoustic mode field profiles, thereby enhance the photon-phonon interaction. Furthermore, we experimentally investigate the impact of varying waveguides direction, defined as the rotation angle $\theta$ with respect to the crystalline $z$-axis, to optimize the photoelastic interaction which directly determines the achievable Brillouin gain.

The pattern for an array of 3\,cm-long meander-shaped LN waveguides is defined through an electron beam lithography (EBL) process with hydrogen silsesquioxane resist~\cite{lu2019}. A second EBL process based on ZEP520A is utilized for patterning the releasing windows. Reactive ion etching (RIE) is employed to etch the patterned LN layer down to the silica layer below. The chip is subsequently immersed in a 10:1 buffered oxide etchant (BOE) solution for 45\,mins to release the underlying silica, thereby achieving the suspended LN waveguide structure. A cross-sectional view of the suspended waveguide is schematically depicted in Fig.\,\ref{devicedesign}(b). The corresponding waveguide dimensions, optimized for the confinement of both optical and acoustic modes through simulations, are designed to be $w=650$\,nm, $\beta=60^\circ$, and $t=190$\,nm. Figure\,\ref{devicedesign}(c) and (d) present the scanning electron microscope (SEM) images of the respective suspended waveguide region for Brillouin interaction and grating coupler for coupling pump and probe light. Figure~\ref{devicedesign}(e) and (f) show the numerically simulated profiles of optical and acoustic waveguide modes of this structure, respectively. The central ridge waveguide supports the low-loss transverse electric-like (TE-like) guided optical mode at $\lambda$ = 1550\,nm (Fig.\,\ref{devicedesign}(e)) and transverse phonon mode at a frequency of 9.05\,GHz (Fig.\,\ref{devicedesign}(f)).

\begin{figure*}
\includegraphics[width=1.0\textwidth]{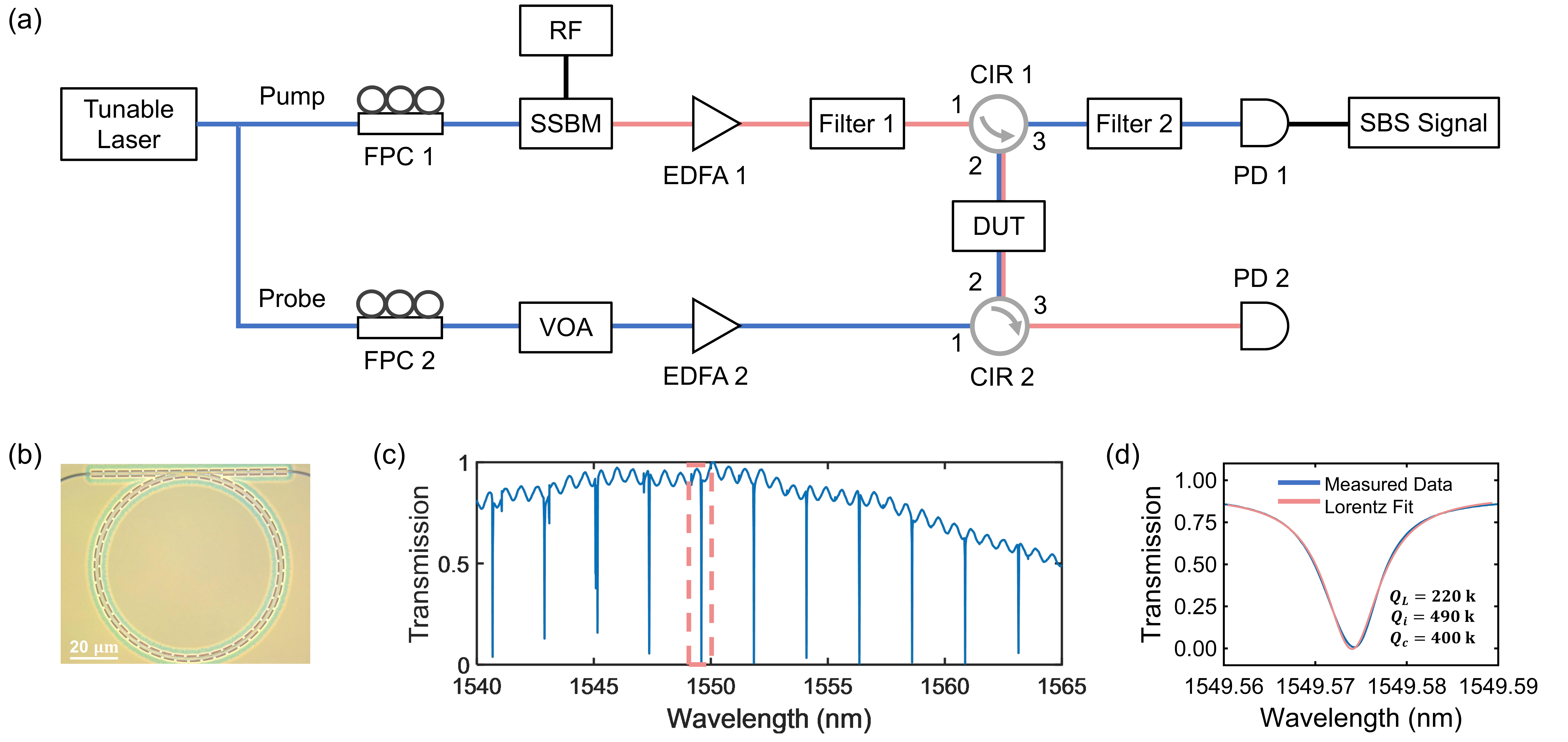}  
\caption{\textbf{Experimental setup and device characterization.} (a) Schematic of the pump-probe experimental setup for the characterization of the Brillouin amplifier. FPC, fiber polarization controller; RF, radio frequency; SSBM, single side band modulator; EDFA, erbium doped fiber amplifier; CIR, circulator; PD, photo-detector; DUT, device under test; VOA, variable optical attenuator. (b) Optical microscope image of a suspended microring resonator with a radius of $70\,\mathrm{\mu m}$ and ring width of 650\,nm. (c) Optical transmission spectrum of the microring resonator for the characterization of the propagation loss. (d) A nearly critical-coupled resonance around 1550\,nm (red dashed box in (c)) with the extracted Q factors.} 
\label{setup}
\end{figure*}

\section{Results and discussion}
The experimental setup, as illustrated in Fig.\,\ref{setup}(a), characterizes the SBS in our devices using a pump-probe scheme. A tunable continuous-wave (CW) laser (Santec 570) is split into two paths to generate the pump light and probe signal, with their polarization controlled to be TE mode using a fiber polarization controller (FPC). In the pump path, the optical wave undergoes a single-sideband modulator (SSBM), resulting in a frequency-upshifted signal. By sweeping the radio frequency (RF) applied to the SSBM, the sideband of the pump will scan across the Brillouin frequency region of the waveguide. The modulated pump light subsequently amplified by a erbium doped fiber amplifier (EDFA 1) and followed by a filter (Filter 1) to suppress the amplified spontaneous emission noise. In the probe path, the laser is amplified by another EDFA 2 and coupled  to the device under test (DUT) in the opposite direction from that of the pump through a grating coupler (Fig.~\ref{devicedesign}(d)). A band-pass filter (Filter 2) is used to block the back-reflected pump light from the signal path, enabling photo-detector (PD 1) to detect the SBS signal, while PD 2 monitors the pump power. This configuration allows for precise measurement of the Brillouin frequency.

With a weak probe power $P_{\mathrm{o}}$ and a pump power $P_{\mathrm{p}}$, the amplified probe power 
\begin{equation}
    P_{\mathrm{S}} = P_{\mathrm{o}}e^{G_{\mathrm{SBS}}L_{\mathrm{eff}}P_{\mathrm{p}}}
    \label{eq:GSBS}
\end{equation} 
is determined by Brillouin gain coefficient $G_{\mathrm{SBS}}$ of backward SBS, with $L_{\mathrm{eff}}=(1-e^{-\alpha L})/\alpha$ representing the effective interaction length and $\alpha$ being the optical propagation attenuation coefficient. To determine the attenuation coefficient, we experimentally fabricated a suspended microring resonator alongside the SBS-generating waveguides using identical dimensional parameters, as shown in Fig.\,\ref{setup}(b). The coefficient $\alpha$ can be derived from the intrinsic quality factor ($Q$) of the microring resonator. Figure\,\ref{setup}(c) depicts a typical transmission spectrum of the suspended microring for a TE-polarized input, where a nearly critical coupled resonance at around 1550\,nm (Fig.\,\ref{setup}(d)) is extracted to have an intrinsic $Q$ of 490k and a corresponding propagation of $\alpha=0.79$\,dB/cm. We further compare the transmittance of suspended meander waveguides with varying lengths, yielding a propagation loss of $\alpha=0.70\pm0.10$\,dB/cm, which aligns well with the value inferred from the intrinsic $Q$. 

\begin{figure}
\includegraphics[width=0.45\textwidth]{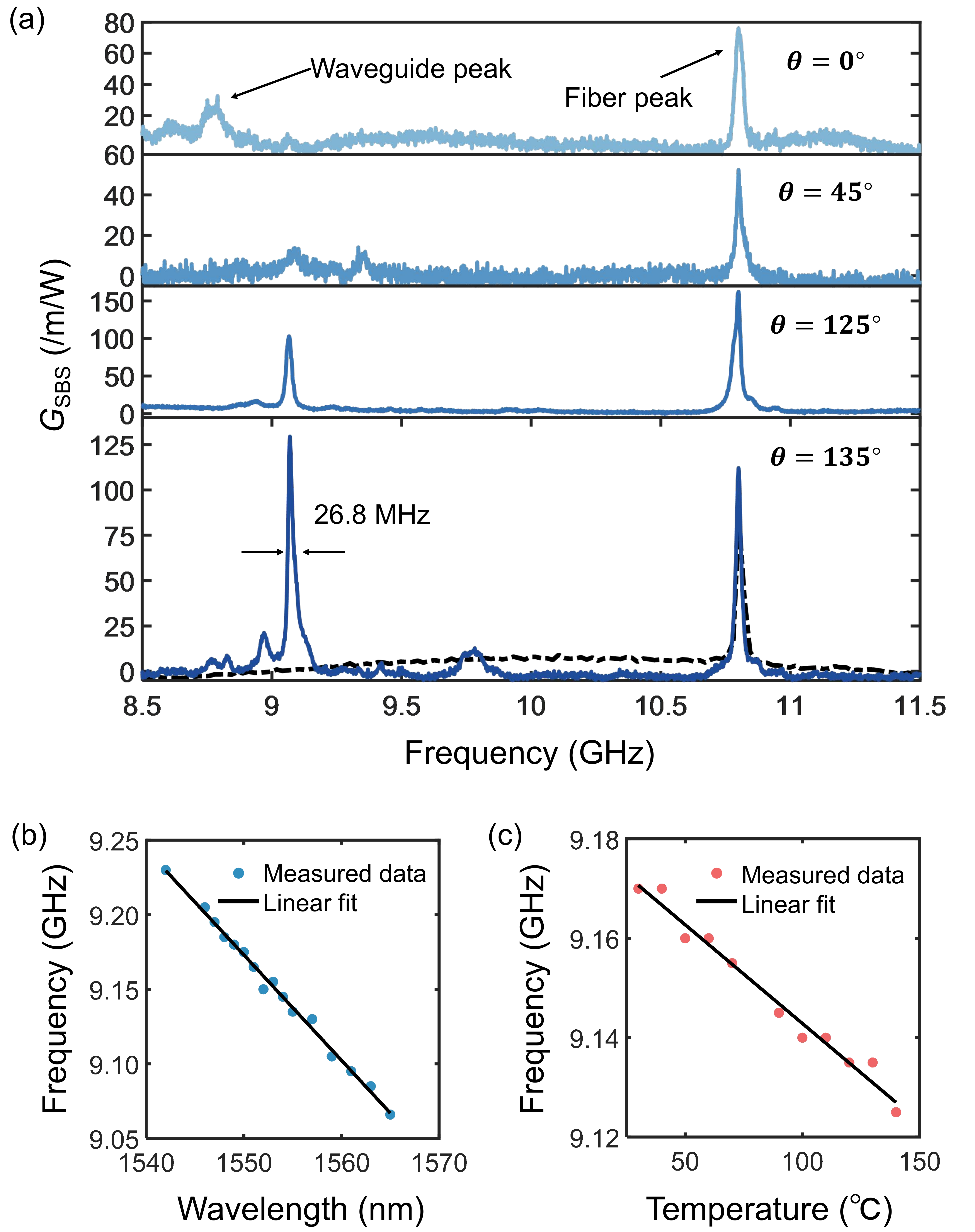}  
\caption{\label{varyangle}\textbf{Angle-dependence of the Brillouin gain coefficient and tuning of Brillouin frequency.} (a) Backward SBS gain coefficient spectra for suspended waveguides with different rotation angles ($\theta$), indicating the significant angle-dependence of $G_{\mathrm{SBS}}$ and a maximum $G_{\mathrm{SBS}}$ of 129.5\,m$^{-1}$W$^{-1}$ is achieved at $\theta=135\,^\circ$. (b) Brillouin frequency versus pump wavelength with a fitted slope of $-7.00\pm0.10$\,MHz/nm, measured at room temperature. (c) Brillouin frequency versus chip operating temperature with a fitted tuning rate of $-0.40\pm0.02$\,MHz/$^\circ\mathrm{C}$,  measured with a pump wavelength of 1550\,nm.} 
\end{figure}

According to Eq.~(\ref{eq:GSBS}), the SBS gain coefficient ($G_{\mathrm{SBS}}$) of the suspended waveguides are calculated. Figure\,\ref{varyangle}(a) depicts the $G_{\mathrm{SBS}}$ for various directions of the waveguide, with rotation angles $\theta=0\,^\circ,45\,^\circ,125\,^\circ$, and $135$\,$^\circ$. Significant discrepancy in the Brillouin gain coefficient is observed for the waveguides with different rotation angles due to the anisotropic photoelastic coefficients of lithium niobate. A maximum Brillouin gain coefficient of 129.5\,m$^{-1}$W$^{-1}$ is achieved at $\theta=135\,^\circ$, which is much higher compared to previously reported values in unsuspended lithium niobate on insulator waveguides~\cite{YeCLEO:24,RodriguesCLEO:24}. For comparison, the SBS response of unsuspened waveguide before releasing the underlying silica layer is characterized and plotted (black dashed line), which shows no detectable SBS signal. This highlights the significant enhancement in the Brillouin interaction provided by the suspended waveguide structure. The measured Brillouin frequency is around 9\,GHz. This value closely matches the predicted frequency of 9.05,GHz for the well-confined phonon mode, as determined by the simulated backward SBS phase-matching condition. We note that there is a slight Brillouin frequency shift in waveguides with the varying rotation angles, which is mainly attributed to the variations in the acoustic velocity of lithium niobate along different crystallographic orientations. Furthermore, the full width at half maximum (FWHM) linewidth of the Brillouin gain in the suspended waveguide with $\theta=135\,^\circ$ is characterized to be 26.8\,MHz, which is 1.5 times narrower than the Brillouin linewidth observed in chalcogenide materials\cite{morrisonCompactBrillouinDevices2017,neijtsOnchipStimulatedBrillouin2024,songStimulatedBrillouinScattering2021}. This narrowband optical amplification is crucial for the development of a narrow-linewidth Brillouin laser. 

The Brillouin frequency $\nu_B$ depends on the both the optical and acoustic properties of the LN waveguides and is given as $\nu_{\mathrm{B}}=2n_{\mathrm{eff}}\mathrm{v}_{\mathrm{a}}/\lambda_{\mathrm{p}}$, where $n_{\mathrm{eff}}$ is the effective index of the optical mode, $\mathrm{v}_{\mathrm{a}}$ is the acoustic velocity of light propagating in the LN waveguides, and $\lambda_{\mathrm{p}}$ is the pump wavelength. With the increasing pump wavelength, the Brillouin frequency exhibits a red-shift with a slope of $-7.00\pm0.10$\,MHz/nm at room temperature, as shown in Fig.\,\ref{varyangle}(b). Since $n_{\mathrm{eff}}$ of LN is temperature-dependent~\cite{morettiTemperatureDependenceThermooptic2005}, we investigate the tuning of Brillouin frequency with the varying temperature with a fixed pump wavelength of 1550\,nm in suspended LN waveguides. As shown in Fig.\,\ref{varyangle}(c), the Brillouin frequency is red-shifted as the temperature increases, exhibiting a slope of $-0.40\pm0.02$\,MHz/$^\circ\mathrm{C}$. This highlights the tunability of the on-chip Brillouin frequency with temperature, paving the way for new possibilities in the development of on-chip sensors.

\begin{figure}
\includegraphics[width=0.45\textwidth]{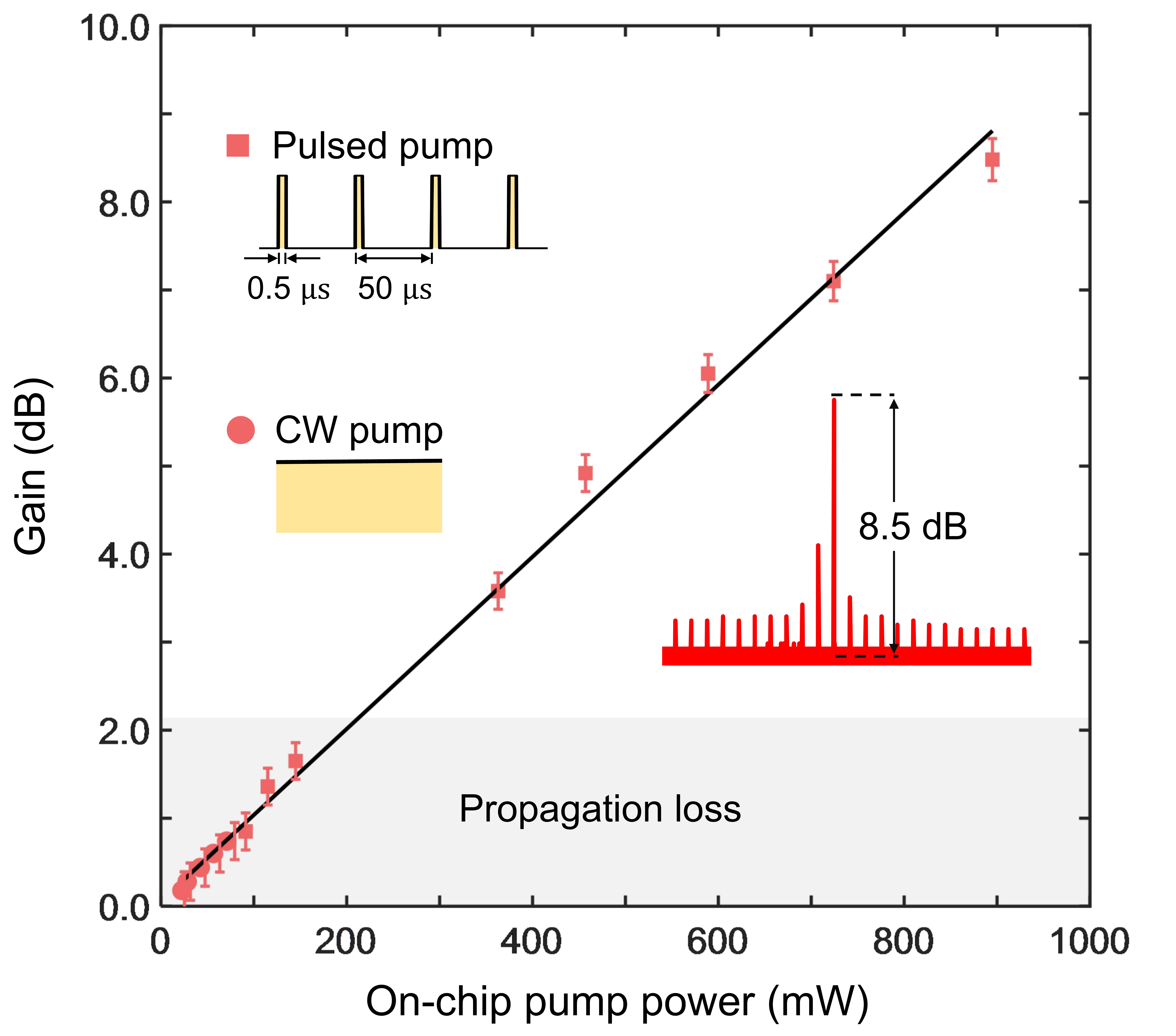}  
\caption{\label{varypower}\textbf{Brillouin amplification.} Brillouin gain as a function of pump power. At low power ($<$90\,mW), CW pump is used, while a modulated pulsed pump is employed at higher power ($>$90\,mW) to reduce the thermal load. The inset illustrates an ultra-high SBS gain of 8.5\,dB, achieved using a pulsed pump with a on-chip peak power of 895\,mW. The error bars represent the variations resulting from the inaccuracy in propagation loss calibration.} 
\end{figure}

To investigate the behavior of devices with respect to the on-chip pump power, we explore it based on the suspended waveguide with $\theta=125\,^\circ$. The Brillouin-interaction-induced gain exhibits a linear increase with the pump power, as shown in Fig.\,\ref{varypower}. A linear fit to the measured gain versus on-chip pump power yields $G_{\mathrm{SBS}}=97.1\pm3.3$\,m$^{-1}$W$^{-1}$, which is in close agreement with the experimentally measured value of 103.0\,m$^{-1}$W$^{-1}$ shown in the third panel of Fig.\,\ref{varyangle}(a). At higher on-chip pump power levels ($>$90\,mW), we switch from CW pump to pulsed pump with duty cycle of 1\%, using an acoustic-optic modulator (AOM) driven by an arbitrary wave generator (AWG) to reduce the thermal load onto the waveguide. A peak on-chip amplification of 8.5 dB is achieved with a pump power of 895\,mW, as depicted in the inset of Fig.\,\ref{varypower} , which is about three times higher than previously reported values in unsuspended lithium niobate on insulator\,\cite{yeBrillouinPhotonicsEngine2024} and represents the highest SBS gain reported on the TFLN platform to date. This increase is primarily due to the suspended structure enhancing both optical and, more notably, acoustic modes confinement. The current maximum gain is limited by the power capacity of our EDFA and the coupling efficiency of the grating couplers. A further increase in pump power is expected to facilitate the observation of higher gain levels. Moreover, the devices performance can be further improved by refining the rotation angle in finer increments to fully leverage the maximum photoelastic coefficient, and by optimizing the fabrication process to reduce the propagation loss of both optical and acoustic modes. Another promising method is to identify an acoustic mode that exhibits a bound state in the continuum (BIC), as an analogue of the photonic BIC in wedge waveguides~\cite{Zou2015,Yu2019}. The BIC mechanism enables much lower radiation loss in a waveguide structure and enhances the SBS interaction, further increasing the Brillouin gain coefficient.

\section{Conclusion}

In conclusion, we have experimentally demonstrated an on-chip Brillouin amplifier based on suspended x-cut LN nanowaveguides. The highest gain coefficient of 129.5\,m$^{-1}$W$^{-1}$ is observed at a rotation angle of 135\,$^\circ$ in our waveguide structure. Meanwhile, Brillouin frequency tuning is investigated by varying the pump frequency and chip temperature, where the respective tuning rates of $-7.0\pm0.1$\,MHz/nm and $-0.40\pm0.02$\,MHz/$^\circ\mathrm{C}$ are realized. By employing a pulsed pump, we are able to demonstrate a new-record high gain for SBS up to 8.5\,dB. This result opens new possibility for RF and photonic signal processing, waveform synthesis, sensing, and narrow-linewidth laser sources.

\vspace{2 mm}
\noindent\textbf{Funding} National Natural Science Foundation of China (62305214, 92265210, and 12293053).

\noindent\textbf{Acknowledgments} The facilities used for device fabrication were supported by the ShanghaiTech Material Device Lab (SMDL). J. Lu acknowledges support from the ShanghaiTech University startup funding. C.L.Z. acknowledges supports from the USTC Center for Micro and Nanoscale Research and Fabrication, and USTC Research Funds of the Double First-Class Initiative.

\noindent\textbf{Disclosures} The authors declare no conflicts of interest.

\noindent\textbf{Data availability} The data that  support the findings of this study are available from the corresponding author upon reasonable request.

%\section{acknowledgments}
%This work is supported by National Natural Science Foundation of China (Grants No.\,62305214, 92265210, and 12293053). The facilities used for device fabrication were supported by the ShanghaiTech Material Device Lab (SMDL). J. Lu acknowledges support from the ShanghaiTech University startup funding. C.L.Z. acknowledges supports from the USTC Center for Micro and Nanoscale Research and Fabrication, and USTC Research Funds of the Double First-Class Initiative. 

%\appendix

%\nocite{*}

%\bibliography{References}% Produces the bibliography via BibTeX.

\end{document}